\documentclass[format=acmsmall, review=false, screen=true]{acmart}
\AtBeginDocument{%
  \providecommand\BibTeX{{%
    \normalfont B\kern-0.5em{\scshape i\kern-0.25em b}\kern-0.8em\TeX}}}

\usepackage{booktabs} 
\usepackage[ruled]{algorithm2e} 

\SetAlFnt{\small}
\SetAlCapFnt{\small}
\SetAlCapNameFnt{\small}
\SetAlCapHSkip{0pt}
\IncMargin{-\parindent}

\usepackage{algorithmic}
\usepackage{graphicx}
\usepackage{textcomp}
\usepackage{xcolor}
\usepackage{multirow}
\usepackage{longtable}
\usepackage{lscape}
\usepackage{algorithmic}
\usepackage{color, colortbl}
\usepackage{rotating}
\usepackage{wrapfig}
\usepackage{tcolorbox}
\usepackage{subcaption}
\usepackage{pdflscape}
\usepackage{hyperref}
\usepackage{wasysym}
\usepackage{booktabs}

\usepackage{amsmath}
\usepackage{array}
\usepackage{hyperref}
\usepackage{soul}

\acmJournal{TOSEM}
\acmVolume{?}
\acmNumber{?}
\acmArticle{39}
\acmYear{2021}
\copyrightyear{2021}
\acmPrice{15.00}

\begin{document}

\setcopyright{acmlicensed}

\acmDOI{0000000.0000000}

\received{XX 2021}
\received[revised]{?? 2021}
\received[accepted]{?? 2021}

\title{Defect Identification, Categorization, and Repair:\newline Better Together}


\author{Chao Ni}
\authornotemark[2]
\affiliation{%
  \institution{School of Software Technology, Zhejiang University}
  \city{Hangzhou}
  \state{Zhejiang}
  \country{China}
  \postcode{310007}
}
\email{chaoni@zju.edu.cn}

\author{Kaiwen Yang}
\authornotemark[2]
\affiliation{%
  \institution{College of Computer Science and Technology, Zhejiang University}
  \city{Hangzhou}
  \country{China}}
\email{kwyang@zju.edu.cn}

\author{Xin Xia}
\authornote{Xin Xia is the corresponding author. \\\ Chao Ni and Kaiwen Yang have equal contribution.}
\affiliation{%
  \institution{Software Engineering Application Technology Lab, Huawei}
  \city{Hangzhou}
  \country{China}
}
\email{xin.xia@acm.org}

\author{David Lo}
\authornotemark[4]
\affiliation{%
  \institution{Singapore Management University}
  \country{Singapore}
}
\email{davidlo@smu.edu.sg}

\author{Xiang Chen}
\authornotemark[3]
\affiliation{%
  \institution{School of Information Science and Technology, Nantong University}
  \city{Nantong}
  \country{China}
}
\email{xchencs@ntu.edu.cn}

\author{Xiaohu Yang}
\authornotemark[2]
\affiliation{
  \institution{Computer Science and Technology,Zhejiang University}
  \city{Hangzhou}
  \country{China}
}
\email{yangxh@zju.edu.cn}

\renewcommand{\shortauthors}{Chao Ni and Kaiwen Yang, et al.}

\begin{abstract}

Just-In-Time defect prediction (JIT-DP) models can identify defect-inducing commits at check-in time.
Even though previous studies have achieved a great progress, these studies still have the following limitations: 
1) useful information (e.g., semantic information and structure information) are not fully used;
2) existing work can only predict a commit as buggy one or clean one without more information about what type of defect it is;
3) a commit may involve changes in many files, which cause difficulty in locating the defect; 
4) prior studies treat defect identification and defect repair as separate tasks, none aims to handle both tasks simultaneously.

In this paper, to handle aforementioned limitations, we propose a comprehensive defect prediction and repair framework named {\sc CompDefect},
which can identify whether a changed function (a more fine-grained level) is defect-prone, categorize the type of defect, and repair such a defect automatically if it falls into several scenarios, e.g., defects with single statement fixes, or those that match a small set of defect templates.
Generally, the first two tasks in {\sc CompDefect} are treated as a multiclass classification task, while the last one is treated as a sequence generation task.
The whole input of {\sc CompDefect} consists of three parts {(exampled with positive functions)}: the clean version of a function (i.e., the version before defect introduced), the buggy version of a function and the fixed version of a function.
In multiclass classification task, {\sc CompDefect} categorizes the type of defect via multiclass classification with the information in both the clean version and the buggy version.
In code sequence generation task, {\sc CompDefect} repairs the defect once identified or keeps it unchanged.

To verify the effectiveness of {\sc CompDefect}, we first build a large-scale function-level dataset named Function-SStuBs4J, which totally contains 21,047 instances with three versions of modified functions and we then evaluate {\sc CompDefect} with state-of-the-art approaches in various settings.
Experimental results indicate the promising performance of {\sc CompDefect} over a set of benchmarks. 
Specifically, for defect identification task, on average,  {\sc CompDefect} improves DeepJIT and CC2Vec by 39.0\% and 41.7\%, by 34.7\% and 37.3\% in terms of F1-score and AUC, respectively.
For defect categorization task, on average, {\sc CompDefect} also improves pre-trained models (i.e., BERT, RoBERTa and CodeBERT) by at least 63.0\%.
For defect repair task, {\sc CompDefect} still outperforms SequenceR and make an improvement by 23.9\% and 29.5\% in terms of BLEU and Accuracy, respectively.

\end{abstract}


\maketitle

\section{Introduction}
\label{introduction}

Software development process is evolving rapidly with frequently changing requirement, various development environment, and diverse application scenarios.
It tends to release software versions in a short-term period.
Such a rapid software development process and limited Software Quality Assurance (SQA) resources have formed a strong contradiction.
Thus, many continuous code quality tools (e.g., CI/CD, static analysis) have been widely adopted~\cite{pornprasit2021jitline}.
However, SQA teams cannot effectively inspect every commit with limited SQA resources.
Therefore, it is important to identify
defects as early as possible to prioritize limited resource (e.g., time and effort) on specific program modules.

Just-In-Time defect prediction (JIT-DP)~\cite{kamei2013large,liu2017code,pornprasit2021jitline,hoang2019deepjit} is a novel technique to predict whether a commit will introduce defects in the future and it can help practitioners prioritize limited SQA resources on the most risky commits during the software development process.
Compared with coarse-grained level (i.e.,
class/file/module) defect prediction approaches,
JIT-DP works at the fine-grained level to provide hints about potential defects.
Though many approaches have been proposed to make a  great process in JIT-DP,
there still has several limitations in previous work.
\begin{itemize}
    \item \textbf{Code semantic information and code structure information are not fully used}. 
Many approaches~\cite{huang2018identifying,McIntoshK18,yang2017tlel} are proposed based on the commit-level metrics proposed by Kamei et al.~\cite{kamei2013large}.
These metrics are quantitative indicators of modified codes without considering the semantic of codes.
Recently, deep learning based approaches~\cite{hoang2019deepjit,hoang2020cc2vec} consider the semantic (i.e., the code tokens' implication around its modified context) and the structure information (e.g., the relation among commits,  $hunk$s\footnote{https://git-scm.com/}, modified files, modified lines, and tokens) of a change.
However, the semantic information is not comprehensive~\cite{hoang2019deepjit} but only represents the modified lines and is part of the understanding of the code commit.
The structure information is not real representation of code structure~\cite{hoang2020cc2vec} but only represents the structure of $\mathit{git\ diff}$.
Therefore, the semantic information (e.g., the code tokens' implication around its modified or unmodified context) and the structure information (e.g., data flow information of code) should be deeply excavated and fully used simultaneously.
\item \textbf{Prediction type is coarse-grained.}
Existing work~\cite{kamei2013large,hoang2019deepjit,huang2017supervised} can only predict whether a commit is buggy or clean  without more information about the  type of defect.
Information about the defect type (e.g., $\mathit{Missing\ Throws\ Exception}$ and $\mathit{Less\ Specific\ If}$) may help developers better understand the categories of defect and consequently help fix it.
\item  \textbf{Location of the defect is not accurate enough}.
Currently, JIT-DP approaches can only predict the defect-proneness of a commit.
However, a commit may involves several hunks which may modify a few functions existing in many files. 
Therefore, it may be unclear where the defect exactly exists if a commit is predicted as buggy one.
\item \textbf{Solution to automatically fix defects once identified is scarcely provided}.
Some approaches~\cite{kamei2013large,pornprasit2021jitline} focus on defect identification, while some approaches~\cite{chen2019sequencer,gazzola2017automatic} focus on defect repair.
However, none of prior works treat defect identification task and defect repair task simultaneously.
\end{itemize}

In this paper, to address these limitations, we propose a comprehensive defect prediction and repair framework named {\sc CompDefect} by building a multi-task deep learning model, which can identify whether {a changed function in a commit} is defect-prone, categorize the type of defect, and repair the defect automatically.
Considering that defect repair is an important but difficult software engineering problem, in this paper, we focus on simple types of defect~\cite{karampatsis2020often,chen2019sequencer,wen2018context,saha2017elixir}, such as defects with single statement fixes, or that match a small set of defect templates. 
Therefore, to make {\sc CompDefect} usable in practical applications, we simplify the usage scenario to single statement defects.
In general, {\sc CompDefect} consists of two stages: an offline learning stage and an online application stage. 
In the offline learning stage, we first build a large-scale function-level  dataset named Function-SStuBs4J by extracting the function body where the modified lines exists in ManySStuBs4J dataset, which is originally collected by Rafael-Michael et al.~\cite{karampatsis2020often} from 1,000 popular open-source Java projects and summarized into 16 defect patterns.
In particular, we need to extract three versions of each function body context where the modified lines exists in a commit:
the clean version, the buggy version, and the fixed version.
The clean version means the version before the defect was introduced,
the buggy version means the version when defect was introduced into the function,
and the fixed version means the version when the defect was fixed.
Then, the three versions are treated as the input of {\sc CompDefect} to complete two main tasks: \textbf{(1) a multiclass classification task} and \textbf{(2) a code sequence generation task}.
The first task can help to identify the buggy function and categorize the type of defect, while the second task can generate the patch to repair the defect.
In online application stage, {for a given commit, we firstly identify the modifications in each $hunk$.
Then, for each modification, we extract the function body where the modification exists.
After that, we extract the corresponding previous function body before the modification occurs.}
Finally, each modified function has two versions of function body (i.e., the current version and the version before the modification introduced), which are fitted into the trained {\sc CompDefect} to predict its defect-proneness and subsequently repair it once identified as defect-inducing one.
{
    Moreover, since {\sc CompDefect} identifies the defect-proneness of each function where the modification occurs in a commit, {\sc CompDefect} can locate the bug in a commit in a more fine-grained way. That is, {\sc CompDefect} locates the defect at hunk-level rather than commit-level.
}

To verify the effectiveness of our proposed model {\sc CompDefect}, we conduct a comprehensive studies on Function-SStuBs4J with state-of-the-art approaches on a few tasks involving: 
defect identification task~\cite{hoang2019deepjit,hoang2020cc2vec}, defect categorization task~\cite{devlin2018bert}, and  defect repair task~\cite{chen2019sequencer}. 
Comparing with several state-of-the-art baselines, the superiority of {\sc CompDefect} is highlighted.
In particular, for defect identification task, on average,  {\sc CompDefect} improves DeepJIT and CC2Vec by 39.0\% and 41.7\%, by 34.7\% and 37.3\% in terms of F1-score and AUC, respectively.
For defect categorization task, on average, {\sc CompDefect} also improves pre-trained models (i.e., BERT, RoBERTa and CodeBERT) by at least 63.0\%.
For defect repair task, {\sc CompDefect} still outperforms SequenceR and make an improvement by 23.9\% and 29.5\% in terms of BLEU and Accuracy, respectively.


In summary, this paper makes the following main contributions:
\begin{itemize}
\item \textbf{Technique.} 
     We propose a comprehensive defect prediction and repair framework named {\sc CompDefect} for function-level software maintenance, which can automatically identify the defect-proneness of a changed function in a commit, categorize the type the defect, and repair defect with appropriate patch by studying the relations among three different versions of the changed method/function.
    {\sc CompDefect} can fully use both the code semantic information (e.g., code tokens' context) and code structure information (e.g., data flow) of the changed function simultaneously and the experimental study also highlights the promising ability on a set of software maintenance tasks: defect identification, defect categorization, and defect repair.
    \item \textbf{Dataset.} 
        We extend and purify a new function-level simple statement dataset named Function-SStuBs4J on the basic of ManySStuBs4J. 
        This dataset can provide more contextual information of modified functions in commits.
        To the best of our knowledge, this is the first function-level multiclass dataset which can provide comprehensive information of a changed function in a commit and its corresponding repaired patch.
   \item \textbf{Replication.} 
    We have also released our replication package including the extended dataset and the source code of {\sc CompDefect}, to facilitate other researchers and practitioners to replicate our work and evaluate their ideas.
    \textbf{This replication package is now publicly available on Zenodo.
    \begin{center}\href{https://zenodo.org/record/5353354\#.YS8iYtMzZhE}{https://zenodo.org/record/5353354\#.YS8iYtMzZhE}\end{center}}
\end{itemize}

The rest of the paper is organized as follows. 
Section~\ref{approach} introduces the details of our proposed model {\sc CompDefect} including the tasks definition and technical details.
Section~\ref{setting} describes the experimental setting involving dataset, baselines, evaluation metrics, experiment setup and research questions we want to investigate.
Following that, the results to research questions and analysis are provided in Section~\ref{results}.
Then, the threats to validity are described in Section~\ref{threats}.
Finally, the related work and conclusion of this paper are subsequently presented in Section~\ref{relatedwork} and Section~\ref{conclusion}, respectively.

\section{CompDefect: automatic software defect identification, categorization and repair}
\label{approach}

\begin{figure*}[!htbp]
    \centerline{
        \includegraphics[width=0.95\textwidth]{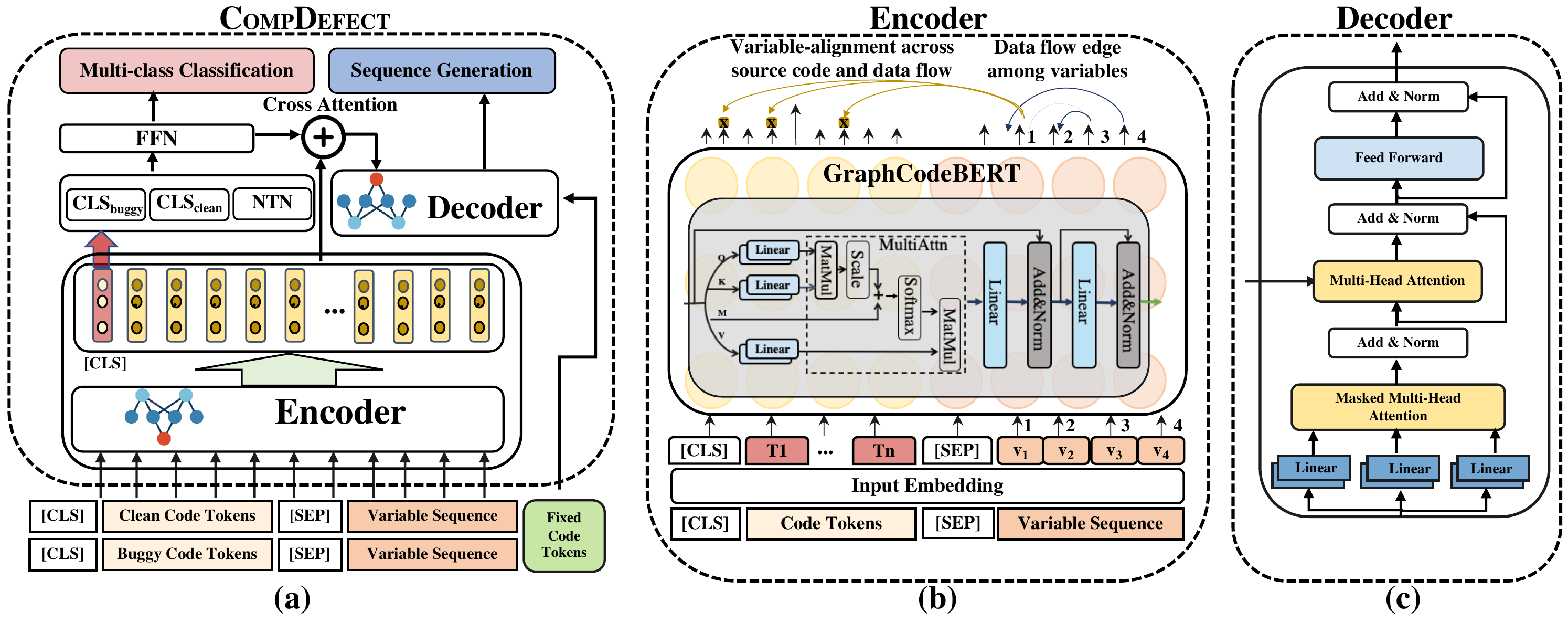}    }
    \caption{{(a) The overall framework of \sc {CompDefect}}. (b) The detail of encoder. (c) The detail of decoder. }
    \label{fig:framework}
\end{figure*}

Multi-task deep learning model~\cite{crawshaw2020multi} aims at training with the same dataset for multiple tasks simultaneously by using the shared representations to learn the common ideas between a few related tasks, which consequently increase efficiency the data and 
potentially accelerate learning speed for related or downstream tasks.
In this paper, our {\sc CompDefect} is a typical multi-task deep learning model, which aims at automatically identifying whether a modification to a function will introduce defect, categorize the type of defect and repairing such a defect by generating patches, simultaneously.
Generally, the first two tasks in {\sc CompDefect} is treated as a multiclass classification task, while the last one is a sequence generation task.
The classification task and the generation task share one common encoder to extract semantic structure information from the function body. When executing the generation task, the knowledge of the classification task is  utilized to help generate a better code repair.

In the rest of this section, we first give the definition of the two tasks.
Subsequently, we introduce details of {\sc CompDefect}, including its encoder and decoder.
We use GraphCodeBERT~\cite{guo2020graphcodebert} as the encoder of {\sc CompDefect} and use Transformer decoder~\cite{vaswani2017attention} as the decoder of {\sc CompDefect}.
For training {\sc CompDefect}, we need three versions of a function as the input.
That is the clean version of a function, the buggy version of a function and the fixed version of a function.
The buggy version represents the state when the defect was introduced into the function,
the clean version represents the state before the defect was introduced, and the fixed version represents the state when the defect was fixed.
{The overall framework of our approach is illustrated in
Fig.~\ref{fig:framework}, which contains three sub-figures.
The left one (i.e., Fig.~\ref{fig:framework}(a)) illustrates the structure of {\sc CompDefect}, while the other two represent the encoder (i.e., Fig.~\ref{fig:framework}(b) is GraphCodeBERT) and decoder (i.e., Fig.~\ref{fig:framework}(c) is Transformer Decoder) used in our proposed {\sc CompDefect}, respectively.
Following that, we define the target multi-task object function since {\sc CompDefect} tries to maximize two objective functions of tasks.
Finally, we illustrate the workflow how {\sc CompDefect} works in the application usage.}

\subsection{Task Definition}
\subsubsection{Defect Identification and Categorization: Multiclass Classification}

The first task of {\sc CompDefect} is to automatically identify defect in each changed function and categorize the type of defect.
To do so, {\sc CompDefect} needs to give the types of defect if it is identified as a defect-prone one.
We formulate this task as a multiclass classification learning problem. 
For a given hunk in a commit, let \textbf{$\mathit{Func}_{\mathit{buggy}}$} be the source code of  buggy version of the specific function where the modified statement exists,
\textbf{$\mathit{Func}_{\mathit{clean}}$} be the source code of clean version of corresponding function before the defect is introduced.

The target is to automatically determine the status $y_i$ of the change between the two versions of the function.
{Let $y_i$ represents the label of the function (i.e., clean or the type of defect) after the modification is introduced. }
In particular, our goal is to build a model $\theta$ with  $\mathit{ \left \langle Func_{clean},Func_{buggy}\right \rangle}$ tuple to maximize the
probability $P_{\theta}\left( y_i \mid	\left \langle \mathit{Func_{clean},Func_{buggy}}\right \rangle; i \in [1, num\_labels] \right)$, over training dataset. Mathematically, our task is defined as finding ${L_1(X)}$, such that:
\begin{gather}
    \mathit{L_1(X)} = argmax_{y_i} \sum_{}^{} log\ P_{\theta}\left( y_i \mid	\left \langle {Func_{clean},Func_{buggy}}\right \rangle; i \in [1, num\_labels] \right)
\end{gather}

$P_{\theta}\left( y_i \mid	\left \langle \mathit{Func_{clean},Func_{buggy}}\right \rangle;  i \in [1, num\_labels] \right)$ can be seen as the conditional likelihood of predicting the status $y_i$ given the $\mathit{ \left \langle Func_{clean},Func_{buggy}\right \rangle}$  input tuple.

\subsubsection{Defect Repair: Sequence Generation}

The second task is to immediately and automatically generate the fixed patch if a changed function is identified as a bug-inducing one.
It is a sequence generation task and the source is buggy function while the target is the corresponding fixed ones, denoted as $Y=\{y_1, y_2, \dots, y_N\}$. 
Let $Y_{j-1}$ be the generated sequence outputted by the decoder of {\sc CompDefect} at position of $\mathit{j}$-$1$, {mathematically $Y_{j-1}=\{y_1, y_2, \dots, y_{j-1}\}$ },
and
$\mathit{X_{Encoder}}$ be the encoded output of the encoder with two inputs (i.e., $\mathit{Func_{clean}}$, $\mathit{Func_{buggy}}$).
Therefore, our task is to maximize the probability $y_{j}^{k}, j \in [1, N],  k \in [1, \mathit{vocab\_size}]$ with $\langle Y_{j-1}, \mathit{X_{Encoder}} \rangle$ tuple over training dataset.
Mathematically, our task is defined as finding $L_2(X)$, such that:

\begin{equation}
    \mathit{L_2(X) = argmax_{y_{j}^{k}}
    \sum_{}^{} log\ {P_{\theta}}\left (y_{j}^{k} \mid \langle Y_{j-1}, X_{Encoder} \rangle; j \in [1,N]; k\in [1, \mathit{vocab\_size}] \right)}
\end{equation}

$P_{\theta}\left (y_{j}^{k} \mid \langle Y_{j-1}, X_{Encoder} \rangle; j \in [1,N]; k\in [1, \mathit{vocab\_size}] \right)$ can be seen as the conditional likelihood of predicting the status $y_{j}^{k}$ given the $\mathit{ \langle Y_{j-1}, X_{Encoder} \rangle }$ input tuple.

\subsection{Classification Task}
As for this task,  we use GraphCodeBERT~\cite{guo2020graphcodebert} as the encoder of {\sc CompDefect}, which is a pre-trained model for programming language considering the structure of code.
In particular, it utilizes the semantic-level information of code (i.e., data flow) for pretraining instead of using syntactic-level structure of code (i.e., abstract syntax tree).
Data flow of a code is represented as a graph, in which nodes represent variables, while edges represent the relation among  variables.
GraphCodeBERT learns a good code representation from source code and its code structure through completing two structure-aware pre-training tasks: data flow edges prediction and variable-alignment across source code and data flow. 
GraphCodeBERT is designed on the basic of Transformer
neural architecture~\cite{vaswani2017attention} by introducing a graph-guided masked attention function to integrate the code structure.

The input of {\sc CompDefect}'s encoder involves two function bodies: clean version of function and buggy version of function.
Then, we use tree sitter\footnote{https://tree-sitter.github.io/tree-sitter/}, a parser generator tool and an incremental parsing library, to transform the two functions into code tokens to build the data flow graph (DFG), which can help to construct the dependencies among variables.
In particular, for a source code $C = \{c_1, c_2,\dots, c_m\}$, 
{\sc CompDefect} first parses the source code into an abstract syntax tree (AST), which includes syntax information of the code.
Besides,  the leaves in AST are used to identify the sequence of variable $ V = \{v_1, v_2, \dots, v_k\}$.
Therefore, the variable is treated as a node of the graph, while the relationship  between $v_i$ and $v_j$ is treated as an directed edge from $v_i$ to $v_j$of graph $e= \left \langle v_i,v_j \right \rangle$, which means the value of $v_j$ comes from $v_i$.
We denote the collection of directed edges as $E=\{e_1, e_2,\dots, e_l\}$ and consequently the graph is denoted as $G(C)=(V,E)$, which represents dependency relationship among variables in the source code $C$.
Then, for the clean version of function, we concatenate clean code and the collection of variables as the input 
sequence $\mathit{X_{clean}=\{[CLS], C_{clean}, [SEP], V_{clean}\}}$ and accordingly, the buggy version of function is transformed as $\mathit{X_{buggy}=\{[CLS], C_{buggy}, [SEP], V_{buggy}\}}$.
$[CLS]$ is a special token in front of the whole sequence, $[SEP]$ is another special token to split two kinds of data types.
After that, we concatenate $\mathit{X_{clean}}$ and $\mathit{{X_{buggy}}}$ vertically to assemble the whole input $X$ for {\sc CompDefect}.

{\sc CompDefect} takes the sequence $X$ as the input and transforms it into input vectors $H^{0}$.
This vector will be transformed for $N$ times, that is, $\mathit{H^{n}=\mathit{transform}_{n}(H^{n-1}), n \in [1,N]}$, in which $N$ represents the maximum number of layers in encoder.
 Finally, we have the semantic structure embedding representation in the last hidden layer,\begin{small} $H^N = \left\{ h_{[CLS]}^N, h_{C}^N,h_{[SEP]}^N, h_{V}^N\right\}$\end{small}, in which $h_{C}^N = \{h_{c_1}^N, h_{c_2}^N,...h_{c_n}^N\}$ and $h_V^{N} = \{h_{v_1}^N,h_{v_2}^N,...,h_{v_n}^N\}$.
 Therefore, we get the contextualized representation of the buggy function $\mathit{H^N_{buggy}}$ and the clean function $\mathit{H^N_{clean}}$.

To learn the relation between the vector of the clean function $\mathit{h_{[CLS]_{clean}}^N}$ and the vector of buggy function $\mathit{h_{[CLS]_{buggy}}^{N}}$, we adopt neural tensor network (i.e., NTN, denoted as $\Gamma$) and denote the relation as $\mathit{h_{NT}}$.
The three vectors are combined as $\mathit{h_{change}}$, which be fed into feed forward neural network for feature fusion.
{\sc CompDefect} finally outputs the probability of each class after a $\mathit{softmax}$ operating on the result of $\mathit{tanh}$ activation function.
More precisely, the classification operation sequence is formulated as follows:
 \begin{gather}
h_{NT}=\operatorname{ReLU}\left( { \left (h_{[\mathit{CLS}]_{\mathit{buggy}}}^{N} \right )}^\mathsf{T} \Gamma^{[1, \ldots, n]}
{h}_{[\mathit{CLS}]_{\mathit{clean}}}^{N}+b_{\mathit{NT}}\right)\\
h_{\mathit{change}} = h_{\mathit{NT}} \oplus h_{[\mathit{CLS}]_{\mathit{buggy}}}^{N} \oplus h_{[\mathit{CLS}]_{\mathit{clean}}}^N\\
e = \operatorname{\mathit{tanh}}\left (\operatorname{FFN}\left (h_{\mathit{change}}\right ) \right) = \operatorname{tanh}\left ({W_e}h_{\mathit{change}} + b_e \right )\label{equtation:e}\\
y = \operatorname{Softmax}(p)= \operatorname{Softmax} (\operatorname{FFN(e)}) = \operatorname{Softmax}( W_pe + b_e)
\end{gather}

To train a good model for our task, we reuse the pre-trained weights of GraphCodeBERT from Hugging Face\footnote{https://huggingface.co/} to initialize the encoder and fine-tune on our collected dataset.

\subsection{Sequence Generation Task}

As for this task,  we use Transformer decoder~\cite{vaswani2017attention} as the decoder of {\sc CompDefect}, which can generate fix patch for the identified buggy function.
The decoder of {\sc CompDefect} consists of two phases: training phase and generating phase.
In the former phase, the decoder starts from special token [SOS] to generate fixed code tokens according to the language model.
For the result in $j$-th position $y_j$, the input of decoder is the output of {\sc CompDefect}'s encoder and the output generated by decoder in (${j}$-1)-th position, denoted as 
$Y_{j-1} = \left\{y_1, y_2, \dots, y_{j-1}\right\}$.
$Y_{j-1}$ is converted into $H_j^0$ and subsequently transformed into $H_j^n = \mathit{transfom_n}(H_j^{n-1}), n \in [1,N]$ by the decoder network.
$N$ is the maximum number of layers in decoder.
Following that, we use $\mathit{softmax}$ function to convert the output values based on last hidden state into probabilities to choose the most likely token from a vocabulary. More precisely,
\begin{gather}
e_j = \mathit{tanh}\left(\mathit{FFN}\left(H_j^N\right)\right) = \mathit{tanh}\left({W}H_{j}^N + b \right)\\
p_j = \mathit{FFN\left(e_j\right)} = We_j + b\\
y_j = \mathit{softmax}\left({p_j}\right)
\end{gather}

in which, $\mathit{FFN}$ represents a feed-forward network and $\mathit{tanh}$ is used as the activation function.
To fully use the information embedded in encoder, we use a cross multi-head attention vector $e_{\mathit{cross}}$:
\begin{gather}
    e_{\mathit{cross}}=e \oplus h_{C_{\mathit{buggy}}}^{N} \oplus h_{[\mathit{SEP}]_{{\mathit{buggy}}}}^{N} \oplus h_{V_{\mathit{buggy}}}^{N}
\end{gather}

Notice that we use $e$, the vector in the encoder phase of {\sc CompDefect} before $\mathit{softmax}$ operation in Equation~(\ref{equtation:e}), instead of $h_{{[\mathit{CLS}]}_{\mathit{buggy}}}^{N}$ to except this attention to focus on the knowledge of classification tasks when performing generation tasks.

In the generation phase, it can be used to generate patches on testing dataset.
In the process of generation, for generating $y_j$, we use beam search strategy to generate multiple potential patches $Y_{j-1}$, as done in prior work~\cite{chen2019sequencer,tufano2019empirical}.
Beam search works through memorizing the $n$ best sequences up to the current state of decoder.
$n$, set as 10 in our study, is commonly referred to as width or beam size, and an infinite $n$ beam search corresponds to a complete breath-first-search.
The successors of these memorized states are computed and sorted based on their cumulative probability.
Then, the next $n$ best sequences are passed to the next state of decoder.
When evaluating {\sc CompDefect}, we only choose the sequence with the highest probability.

\subsection{Optimization of {\sc CompDefect}}

When a model has more than one task, a few task-specific objective functions need to be combined into a single aggregated one that the model tries to maximize it. 
Therefore, it is extremely important to exactly combine various objective functions into one that is the most suitable for multi-task learning. 
In {\sc CompDefect}, there mainly exists two tasks: multiclass classification task and sequence generation task.
Since in our usage scenario (i.e., function-level software defect prediction and defect repair), we think the two tasks are equivalently important.
Thus, {\sc CompDefect} addresses the multi-task optimization by balancing the individual objective functions for the two different tasks.
{That is, the final optimization function($L_3(X)$) is the sum of two individual objective functions($L_1(X)$ for classification task, and $L_2(X)$ for sequence generation task), which is the one that {\sc CompDefect} tries to maximize.
Formally,}
\begin{gather}
    L_3(X) = L_1(X) + L_2(X)
\end{gather}

Notice that our model {\sc CompDefect} is flexible, we can optimize our model based on how developers perceive the importance of the two tasks by changing the weights of the objective functions.

{
\subsection{The Workflow of {\sc CompDefect} in Application Stage}
}

\begin{figure*}[htbp]
    \centerline{
        \includegraphics[width=1\textwidth]{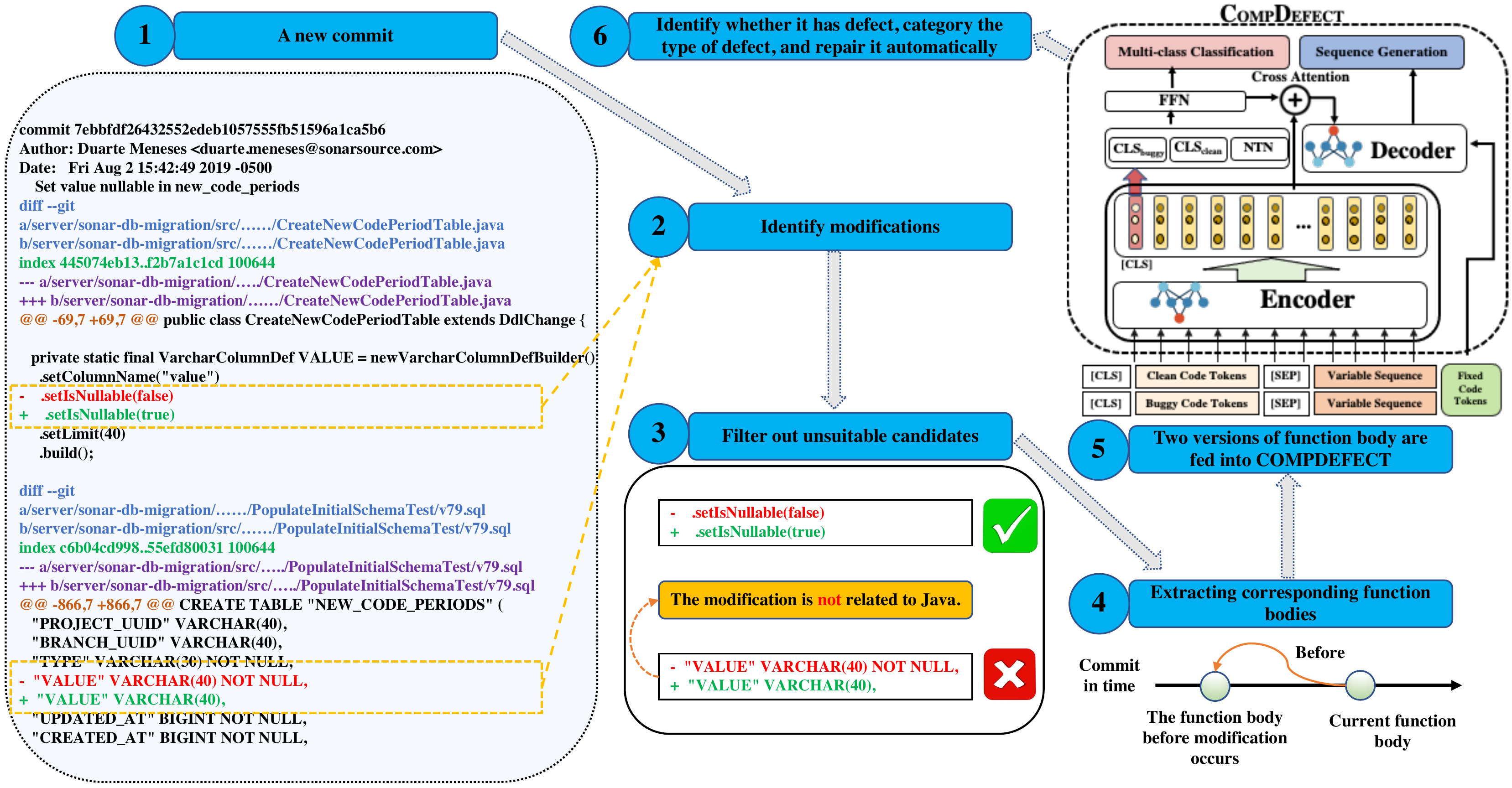}
        }
    \caption{The workflow of {\sc CompDefect} in application stage.}
    \label{fig:applicationworkflow}
\end{figure*}

{
We illustrate how {\sc CompDefect} can be used to process a new commit in Fig.~\ref{fig:applicationworkflow}.
The figure provides an example of a commit(7ebbfdf26432552edeb1057555fb51596a1ca5b6) from $sonarqube$ project.
Given a commit, {\sc CompDefect} firstly identifies hunks. 
Then, it figures out the corresponding function body where the modification exists.
After that, it filters out those hunks whose modification is not a single-statement one or the modification occurs outside the scope of a function.
Following that, {\sc CompDefect} extracts the previous function body of the current version using PyDriller~\cite{Spadini2018}. 
Finally, the two versions are fed into {\sc CompDefect} to judge whether it has a defect, to categorize the type of defect, and to repair it automatically when it is identified as the defective one.
}

\section{Experimental Setting}
\label{setting}

We first introduce the dataset we experiment on, then we present the baselines for different type of tasks.
Following that, the evaluation metrics, experimental 
settings, and research questions are presented subsequently.

\subsection{Dataset}
\label{dataset}
Program repair is an extremely important but difficult software engineering problem.
Fixing simple defects (e.g., one-line defects or defects that fall into a small set of templates) is a good way to obtain acceptable performance.
ManySStuBs4J~\cite{karampatsis2020often} is a collection of single-statement defect-fix change and have two versions: small size version and large size version.
The small version contains 25,539 single-statement defect-fix changes mined from 100 popular open-source Java Maven projects, while the large version contains
153,652 single-statement defect-fix changes mined from 1,000 popular open-source Java projects.
This dataset is annotated by whether those changes match any of a set of 16 defect patterns that appear often.
To make our paper self-contained, we provide a brief introduction of each pattern as follows.
\begin{itemize}
    \item \textbf{Change Identifier Used}. 
        It checks whether an identifier that appears in an expression in a statement is replaced by another.
        It is easy for developers to accidentally use different identifiers instead of having the same type of expected identifier.
        Copying and pasting code is a potential source of such errors.
    \item \textbf{Change Numeric Literal}.
        It checks whether one numeric literal is replaced by another one. 
        It is easy for developers to mix two numeric values in a program.
    \item \textbf{Change Boolean Literal}. 
        It checks whether a Boolean literal was replaced. 
        That is, True is replaced with False and vice-versa.
    \item \textbf{Change Modifier}.
        It checks whether a variable, function, or class is declared with the wrong modifiers. 
    \item \textbf{Wrong Function Name}.
        It checks whether the function was incorrectly called. 
        Functions with similar names and the similar signature are usual pitfall for developers.
    \item \textbf{Same Function More Args}.
        It checks whether an overloaded version of a function with more parameters is called. 
        Functions with multiple overloads often make it confusing to developers.
    \item \textbf{Same Function Less Args}.
        It checks whether an overloaded version
        of the function with less arguments is called. 
        For example, a developer may call a function with at least one default parameter and forget to initialize the parameter.
    \item \textbf{Same Function Change Caller}.
        It checks whether the caller object in the function call expression is replaced by another object. 
        When there are multiple variables having the same type, the developer may perform an operation unexpectedly.
    \item \textbf{Same Function Swap Args}.
        It checks whether a function is called with some of its parameters swapped. When multiple function parameters are of the same type, developers can easily swap two of them without realizing if they do not accurately remember what each argument represents.
    \item \textbf{Change Binary Operator}.
        It checks whether a binary operator is replaced with another one of the same type by accident. 
        For example, developers may often mix up comparison operators in expressions. 
    \item \textbf{Change Unary Operator}.
        It checks whether a unary operator is replaced with another operator of the same type by accident. 
        For example, developers may often forget ! operator in boolean expressions.
    \item \textbf{Change Operand}.
        It checks whether one of the operands in a binary operation is wrong. 
    \item \textbf{More Specific If}.
        It checks whether an additional condition (\&\& operand) is added to the condition of an if statement.
    \item \textbf{Less Specific If}.
        It checks whether an additional condition which either itself or the original one needs to hold ($||$ operand) is added to the condition of the if statement.
    \item \textbf{Missing Throws Exception}.
        It checks whether the defect-fix adds a throw clause in a function declaration.
    \item \textbf{Delete Throws Exception}.
        It checks whether the defect-fix deletes a throw
        clause in a function declaration.
\end{itemize}

However, the details of this dataset is too simple to satisfy our task's requirement.
That is, the original dataset mainly contains the types of defect, {the one statement buggy code and one statement fixed code.}
However, our model needs the whole information about a changed function to capture more information inside the code.
Besides, our study aims at defect identification, we also need to collect negative commits from studied projects since the original dataset only contains the positive commits.
Therefore, we need an extension version of the large size of ManySStuBs4J using the improved toolkit\footnote{https://github.com/h4iku/repairSStuBs}, which fixes the issues in the process of original dataset extraction.

\begin{figure*}[!htbp]
    \centerline{
        \includegraphics[width=1.05\textwidth]{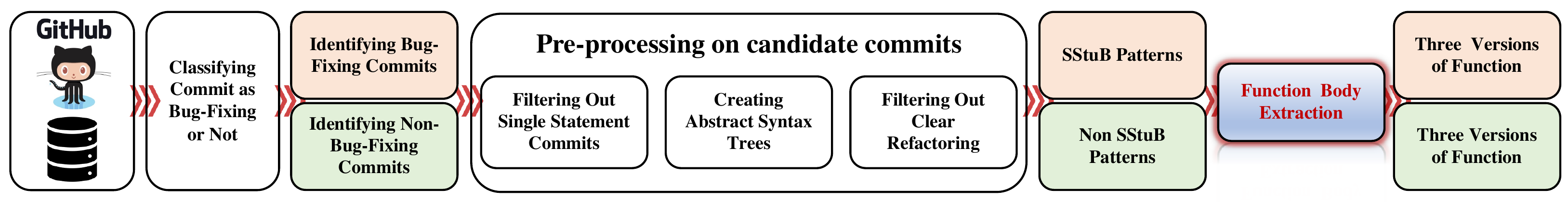}}
    \caption{The process of single statement commits filtering and function body extraction.}
    \label{fig:dataextractprocess}
\end{figure*}

\begin{figure*}[hbtp]
    \centerline{
        \includegraphics[width=1\textwidth]{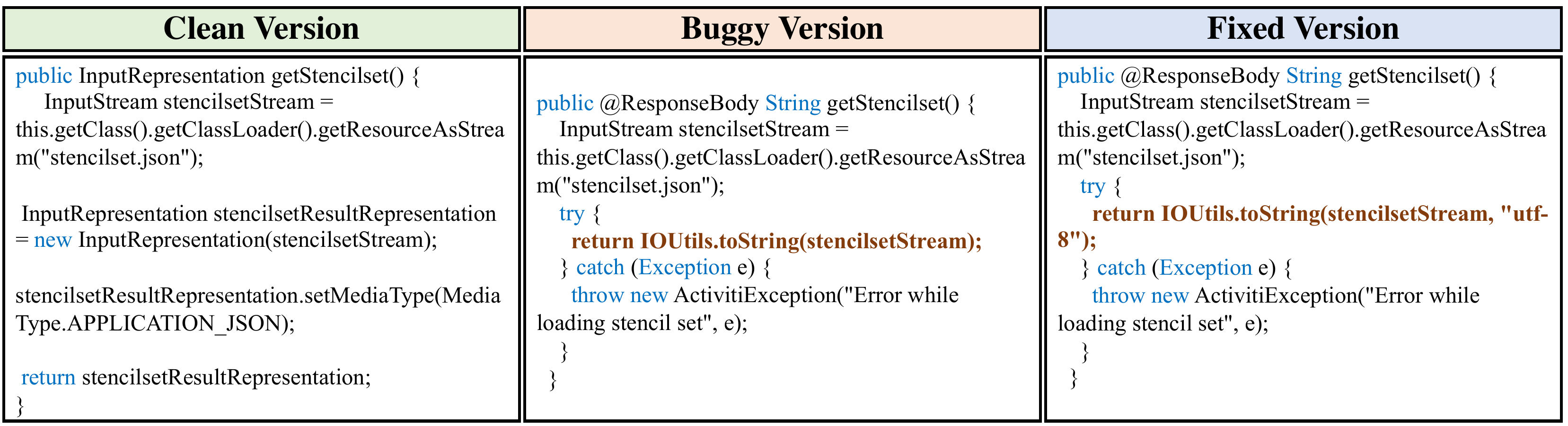}
    }
    \caption{The different contents of a specific function in different states: clean version, buggy version and fixed version.}
    \label{fig:example}
\end{figure*}

Firstly, we take ManySStuBs4J as the basis of the new dataset.
We need more information of changed function in a commit to extend the dataset from two aspects since we try to combine the defect identification task, defect categorization task and defect repair task together:
{
1) extract the function body where the defect-inducing statement or the defect-fix statement exists; 
2) extract the clean function body before the defect-inducing statement was introduced.
}

{
Secondly, the data extraction and extension process is illustrated in Fig.~\ref{fig:dataextractprocess}.   
Notice that we start from the COMMIT rather than the FUNCTION.
In particular, as for extracting the positive/negative functions, 
we follow the same steps (e.g., selecting the same Java projects, identifying non-defect-fixing/defect-fixing commits, selecting single statement changes, creating Abstract Syntax Trees and filtering out clear refactoring) as the authors of ManySStuBs4J~\cite{karampatsis2020often} took: 
\begin{itemize}
\item \textbf{Step 1}. We identify bug-fixing(non-bug-fixing) commits whose commit message contains one(none) of these keywords (i.e., error, buy, fix, issue, mistake, incorrect, fault, defect, flaw, and type).  
That is, we use the opposite selection strategy between selecting positive commits and negative commits. 
\item \textbf{Step 2}. We follow the same strict criteria using in ManySStuBs4J~\cite{karampatsis2020often} to select the commits with single statement modification.
\item \textbf{Step 3}. We identify each scope of the modification (i.e., hunk in git) in the filtered commits and extract their corresponding bug-fixing(non-bug-fixing) function bodies.  That is, we split one commit into hunks and extract the function body of modification in each hunk.
\item \textbf{Step 4}. The SZZ algorithm is used to identify their corresponding bug-inducing(non-bug-inducing) commits. Right now, we can extract the corresponding bug-inducing(non-bug-inducing) function bodies.
\item \textbf{Step 5}. We can extract the corresponding clean versions (i.e., the last version in time before the bug-inducing(non-bug-inducing) functions are modified) using PyDriller. 
Therefore, we can obtain the three versions of the function body for positive/negative functions.
\end{itemize}
}

{
Meanwhile, we also design four criteria for filtering unsuitable functions in (Steps 3-5), such as: }
1) Modified statements exist outside a function.      
This work focuses on the simple scenario, that is, function-level single statement defect prediction and defect repair.  
Therefore, we filter out those statements that lie outside the scope of a function, for example, statements for defining a global variable or object in a class.
2) Defects are introduced in newly added files or functions.
{\sc CompDefect} needs three versions of a specific function.
For newly added defective one, the clean version does not exist.
Therefore, we filter out such cases.
3) Function names cannot be identified. 
On one hand, the modifications in a commit are made to the function name and then it is difficult to solve. On the other hand, the line of function is mapped incorrectly. In ManySStuBs4J, the authors label the line number of buggy or fixed code based on the result of AST, which may not be exactly correct with the original line in source code file.
4) Other failed issues. 
When errors occur in PyDriller or in original dataset, we cannot get the correct result.

\begin{figure}[hbtp]
    \centerline{
        \includegraphics[width=1\textwidth]{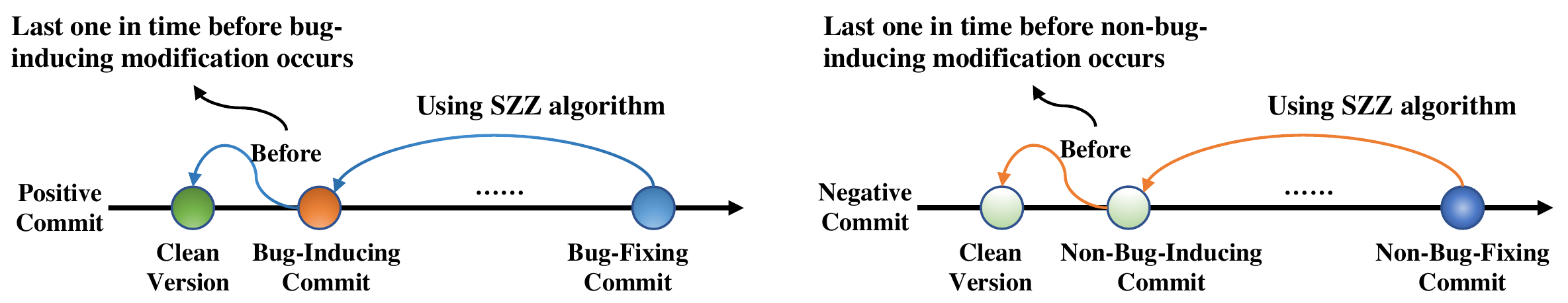}
    }
    \caption{The extraction process of the three versions of function and their relationship.}
    \label{fig:functionextractprocess}
\end{figure}

Take an example from the project of Activiti\footnote{This Activiti is a light-weight workflow and Business Process Management (BPM) Platform targeted at business people, developers and system admins.}  shown in Fig.~\ref{fig:example}.
There exists a function named $\mathit{getStencilset}$\footnote{This function is located at ``
modules/activiti-modeler/src/main/java/org/activiti/rest/editor/main/StencilsetRestResource.java''.
This fixed version is submitted at ``c015d11303339f50254a10be7335fd33546911ab'', while the buggy version is introduced at ``159d1ef8e0cf059165b17bb546f47f639559dfa9''.}.
We extract three versions of this function: clean version, buggy version and fixed version.
Buggy version means the state when the defect-inducing statement is introduced,
clean version means the state before the defect-inducing statement is introduced,
and fixed version means the state when the defect is fixed.
The relationship among the three version can be illustrated in Fig.~\ref{fig:functionextractprocess}.
We extract the three versions of function with the help of PyDriller~\cite{Spadini2018}, which is a Python framework that helps developers on mining software repositories.

Finally, we build the dataset and name it as \textbf{Function-SStuBs4J} for defect identification, defect categorization and defect repair and the statistical information is shown in Table~\ref{tab:dataset}.

\begin{table}[htbp]
    \centering
    \caption{The statistics for Function-SStuBs4J}
    \scalebox{.8}{
      \begin{tabular}{|l|c|c||c|c|c|}
      \hline
      \textbf{Defect Type} & \textbf{Count} & \textbf{Ratio} & \multicolumn{1}{l|}{\textbf{Defect Type}} & \textbf{Count} & \textbf{Ratio} \\
      \hline \hline
      Same Function Change Caller & 381   & 1.81\%  & \multicolumn{1}{l|}{Same Function More Args} & 441   & 2.1\% \\
      \hline
      Change Identifier Used & 1,599  & 7.6\%   & \multicolumn{1}{l|}{Same Function More Args} & 1,169  & 5.55\% \\
      \hline
      Change Modifier & 329   & 1.56\%  & \multicolumn{1}{l|}{Same Function Swap Args} & 139   & 0.66\% \\
      \hline
      Change Numeric Literal & 991   & 4.71\%  & \multicolumn{1}{l|}{Change Boolean Literal} & 330   & 1.57\% \\
      \hline
      Change Operand & 161   & 0.76\%  & \multicolumn{1}{l|}{Less Specific If} & 484   & 2.3\% \\
      \hline
      Change Binary Operator & 523   & 2.48\%  & \multicolumn{1}{l|}{More Specific If} & 570   & 2.71\% \\
      \hline
      Change Unary Operator & 338   & 1.61\%  & \multicolumn{1}{l|}{Missing Throws Exception} & 10    & 0.05\% \\
      \hline
      Wrong Function Name & 3,012  & 14.31\% & \multicolumn{1}{l|}{Delete Throws Exception} & 32    & 0.15\% \\
      \hline
      \multicolumn{1}{|c|}{\textbf{ALL}} & \multicolumn{5}{c|}{\textbf{21,047}} \\
      \hline
      \multicolumn{1}{|c|}{\textbf{Positive}} & \multicolumn{5}{c|}{\textbf{10,509}} \\
      \hline
      \multicolumn{1}{|c|}{\textbf{Negative}} & \multicolumn{5}{c|}{\textbf{10,538}} \\
      \hline
      \end{tabular}%
    }
    \label{tab:dataset}
  \end{table}%

\subsection{Baselines}
\label{baseline}

Since there has no existing work which can identify whether a modification to a function may introduce defects, subsequently categorize the type of defect, and consequently repair it automatically, we make a comprehensive comparison among three types of baselines and {\sc CompDefect}: baselines for defect identification, baselines for defect categorization and baselines for program repair.

For the defect identification task, we consider two well-known methods as baselines: DeepJIT~\cite{hoang2019deepjit} and CC2Vec~\cite{hoang2020cc2vec} since they are deep-learning based methods using the textual information of code.
We briefly introduce the two methods as follows.

\noindent
$\blacksquare\ $\textbf{DeepJIT} is an end-to-end deep learning framework for Just-in-Time defect prediction. 
DeepJIT adopts a Convolutional Neural Network (CNN) model to automatically learn high-dimensional semantic features for commits.
    In particular, DeepJIT uses two CNN models to learn the representation of two parts of the input: one CNN for the commit message and another one for the code commits.
    Finally, the concatenation of two representations is treated as the input of fully-connected layer to output the probability
    of defect-introducing commit.

\noindent
$\blacksquare\ $\textbf{CC2Vec} is a distributed representation learning framework of commit. 
    CC2Vec believes that the commit has hierarchical structure, which is ignored by DeepJIT.
    In particular, one commit is composed of a few changed files,
    one changed file is composed of a few hunks\footnote{https://git-scm.com/}, one hunk is composed of a few changed lines and one changed line is composed of a few changed tokens.
    To grasp the information of hierarchical structure in commits,
    CC2Vec models the Hierarchical Attention Network (HAN) with the help of the attention mechanism and multiple comparison functions are used to identify the difference of modified code (i.e., added code and removed code).
    CC2Vec targets at learning a representation of code commits guided by their accompanying commit messages, which represents the semantic intent of
    the code commits. 

For defect categorization task, we choose three methods BERT~\cite{devlin2018bert}, RoBERTa~\cite{liu2019roberta} and CodeBERT~\cite{feng2020codebert} as the baselines and briefly introduce them as follows.

\noindent
$\blacksquare\ $\textbf{BERT} is designed to pre-train deep bidirectional representations from unlabeled text by jointly conditioning on both
        left and right context in all layers and its capability of capturing the semantic and context information of sentence has been verified in many work~\cite{gao21todo,lewis2019bart,zhang2020sentiment}.       
        BERT consists of 12-layer transformers and each of the transformers is composed of a self-attention sub-layer with multiple attention heads. 
        BERT takes sequence of tokens as the input of embedding component.
        Given a sequence of tokens $x=\{x_1, ..., x_T\}$ with length of $T$,
        BERT will calculate the contextualized representations
        $H^l= \{h_{1}^{l}, ..., h_{T}^{l}\} \in  \Re ^{T \times D}$, where $l$ represents the $l_{th}$ transformer layer and $D$ represents the dimension of the representation vector. 
        As a result, the pre-trained BERT model can be fine-tuned with just one additional output layer to create state-of-the-art models for a wide
        range of tasks, such as binary or multiclass classification.

\noindent
$\blacksquare\ $\textbf{RoBERTa} is a robustly optimized version of BERT by iterating on BERT's pretraining procedure (e.g., training the model longer, using bigger batches over more data) and {\sc CompDefect}'s encoder has the same architecture with RoBEATa.

\noindent
\textbf{CodeBERT} is another pre-trained model on the basic of BERT for programming language, which is a multi-programming-lingual model pre-trained on NL(Natural Language)-PL(Programming Language) pairs in six programming languages (i.e., Python, Java, JavaScript, PHP, Ruby and Go).

For defect repair task, we also consider one state-of-the-art method SequenceR~\cite{chen2019sequencer} as baseline.
We briefly introduce it as follows.

\noindent
$\blacksquare\ $\textbf{SequenceR} is a sequence-to-sequence deep learning model which aims at automatically fixing defects by generating one-line patches.
        The one-line patch means that the defect can be fixed by replacing a single buggy line with a single correct line. 
        SequenceR proposes a novel buggy context abstraction process to organize the fault localization information into a representation.
        Such a representation preserves concise, suitable and valuable information for deep learning model understanding the context of the bug and consequently be used to predict the fix. 
        Then, the representation is  fed to a trained sequence-to-sequence model to execute patch inference.
        In particular, it generates multiple single-lines of code
        that will be treated as the potential one-line patches for the defect.
        Finally, SequenceR generates the most suitable patches by formatting the code and replacing the buggy line with the proposed lines.

\subsection{Evaluation Metrics}

There are four statistics with respect to classification task:
(i) True Positive (TP) represents the number of functions classified as defective and they are truly defective ones. 
(ii) True Negative (TN) represents the number of functions classified as non-defective and they are truly non-defective ones.
(iii) False Positive (FP) represents the number of functions classified as defective  and they are truly non-defective ones. 
(iv) False Negative (FN) represents the number of non-defective functions and they are truly defective ones. 
Therefore, based on the four possible statistics, three widely used performance measures (i.e., $Precision$, $Recall$ and $F1$-$score$) can be defined to evaluate the performance of {\sc CompDefect} and baselines as follows:

\noindent
$\blacksquare\ $\textbf{Precision}: the proportion of functions that are correctly classified as defective among those labeled as defective: $
Precision=\frac{TP}{TP+FP}$.

\noindent
$\blacksquare\ $\textbf{Recall}: the proportion of defective functions that are correctly classified: 
$ Recall=\frac{TP}{TP+FN} $.

\noindent
$\blacksquare\ $\textbf{F1-score}: the harmonic mean of precision and recall.
It evaluates if an increase in precision (or recall) outweighs a reduction in recall (or precision), respectively:
$
F1\text{-}score=\frac{2 \times Precision \times Recall}{Precision + Recall}
$.

\noindent
$\blacksquare\ $\textbf{AUC}: the Area Under the receiver operator characteristics Curve (AUC) is also used to measure the discriminatory power of {\sc CompDefect} and baselines, i.e., the ability to differentiate between defective or non-defective functions. 
AUC calculates the area under the curve plotting the true positive rate (TPR) versus the false positive rate (FPR), while applying multiple thresholds to determine if a function is defect-inducing or not. 
The value of AUC ranges between 0 (the worst discrimination) and 1 (the perfect discrimination).

Besides, to better evaluate how approaches perform on defect repair task, we consider two widely used metrics: BLEU (Bilingual Evaluation Understudy)~\cite{papineni2002bleu} and Accuracy~\cite{chen2019sequencer}.

\noindent
$\blacksquare\ $\textbf{BLEU}: it is a widely used measure for neural machine translation task~\cite{klein2017opennmt} and software related task~\cite{gu2016deep,jiang2017automatically}. 
In our task, it is used to calculate the similarity between the generated code snippet and the referenced correct code snippet.
It has a score range of 0 and 1.
The higher the BLEU score, the closer the generated code is to the referenced one.
It first computes the geometric average of the modified $n$-gram precisions (i.e., $p_n$) by using $n$-grams up to the maximum number of grams $N$ (i.e., set as 4 in our paper) and positive weights $w_n$ summing to one.
Then,  it computes the brevity penalty BP,
$
       BP=\begin{cases}
        1,& if\ c>r \\ 
        e^{(1-r/c)}, & if\ c\le r   
       \end{cases}
$
,in which,  $c$ represents the length of the candidate code snippet and $r$ represents the effective reference corpus length.
Finally, the BLEU score can be calculated as follows:
$
    BLEU = BP\cdot exp\left ( \sum_{n=1}^{N}w_nlog{p_n}  \right ) 
$

\noindent
$\blacksquare\ $\textbf{Accuracy}: the target of defect repair model is to fix as many defects as possible. 
Therefore, we also use accuracy to evaluate the effectiveness of models for defect fixing, it is calculated as the ratio between the number of correctly fixed defects by an approach and the number of total defects to be fixed.

\subsection{Empirical setting.}

We implement our {\sc CompDefect} in Python with the help of Pytorch framework and pre-trained model on Huggingface\footnote{https://huggingface.co/}.
The pre-trained GraphCodeBERT model is used as the encoder for embedding training samples, which can leverage semantic structure of code to learn code representation and also can be easily extended for downstream tasks.
Besides, we also use Transformer decoder~\cite{vaswani2017attention} as the generator of fixed code.
In our model, each version function is embedding as a 768 dimensional vector.
During the training phase, the parameters of {\sc CompDefect} are optimized using Adam with a batch size of 32. 
We also use $ReLu$ and $tanh$ as the activation function.
A dropout of 0.1 is used for dense layers before calculating the final probability. 
The maximum number of epoch in our experiment is 50.
The models (i.e., {\sc CompDefect} and baselines) with the best performance on the validation set is used for our evaluations.

As for dataset split, we use 80\%, 10\% and 10\% of original dataset as training data, validation data and testing data,    respectively.
Notice that, for each part of the data, we keep the distribution among each type of function as same as the original one.

\subsection{Research Questions}

 To comprehensively evaluate the effectiveness of {\sc CompDefect}, we investigate the following research questions. 
\begin{itemize}
    \item\textbf{RQ1:} How does {\sc CompDefect} perform on defect identification compared with state-of-the-art baselines?
    \item\textbf{RQ2:} How does {\sc CompDefect} perform on defect categorization compared with  state-of-the-art baselines?
    \item\textbf{RQ3:} How does {\sc CompDefect} perform on defect repair compared with the state-of-the-art baseline? 
\end{itemize}

\section{Results and Analysis}
\label{results}

\subsection{RQ1: How does {\sc CompDefect} perform on defect identification compared with state-of-the-art baselines?}
\noindent
\textbf{Motivation.}
Just-in-time (JIT) defect prediction has received much attention in the software engineering and many state-of-the-art approaches are proposed~\cite{hoang2019deepjit,hoang2020cc2vec,huang2017supervised}.
These approaches are built from simple model (e.g., CBS+) on manually designed features to complex model (e.g., DeepJIT) on semantic features and have make a great progress in JIT scenario.
As for {\sc CompDefect} based on neural network, it can also identify whether a commit is a defect-inducing one by predicting if the modification to a function will introduce a defect.
Therefore, we want to make a comparison between {\sc CompDefect} with those state-of-the-art semantic features based approaches.

\noindent
\textbf{Approach.}
We treat recently proposed DeepJIT~\cite{hoang2019deepjit} and CC2Vec~\cite{hoang2020cc2vec} as the baseline methods and the introduction of them can be found in Section~\ref{setting}.
There are two differences between {\sc CompDefect} and the two baselines:
1) the two methods can only estimate the defect-proneness of a commit at the commit level and a commit may contain a few hunks which changes a few functions, while {\sc CompDefect} estimates whether a function in a hunk of a commit is defect-inducing. That is, {\sc CompDefect} can estimate the defect-proneness of a commit at the hunk(function)-level, which means {\sc CompDefect} is more fine-grained than the baseline ones.
2) the two baselines can only predict a commit as a defect-inducing one or clean one, while {\sc CompDefect} can not only predict a commit defect-proneness but also can categorize the types of defect.

Considering the two differences and to make a fair commit-level comparison, we make the following two hypothesizes for {\sc CompDefect}: 1) a function in a hunk of a commit will be treated as a buggy one if {\sc CompDefect} predicts it as a non-clean one; 2)  a commit is predicted as defect-inducing one if there exists a least one function in a hunk of a commit predicted by {\sc CompDefect} as the defect-inducing one, otherwise the commit is predicted as a clean one.
Besides, we use the widely used performance measures (i.e.,  Precision, Recall, F1-score, and AUC) to evaluate the difference among those methods.

\noindent
\textbf{Result.}
The evaluation results are reported in Table~\ref{tab:rq1}.
The best performance is highlighted in bold.
According to the results, we find that our approach {\sc CompDefect} has the significant advantage over DeepJIT and CC2Vec on all performance measures.
In particular, {\sc CompDefect} obtains 0.679 and 0.785 in terms of F1-score and AUC, which improves DeepJIT and CC2Vec by 39.0\% and 41.7\%, by 34.7\% and 37.3\% in terms of F1-score and AUC, respectively.
As for Precision and Recall, {\sc CompDefect} also has a large improvement.
Specifically, {\sc CompDefect} improves DeepJIT and CC2Vec by 32.8\% and 38.2\%, by 43.2\% and 44.2\% in terms of Precision and Recall, respectively.
Besides, we surprisingly find that in our scenario, CC2Vec performs a little worse than DeepJIT, which, to some extent, means CC2Vec cannot capture more information than DeepJIT for existing state-of-the-art technique can utilize.

\begin{table}[htbp]
  \centering
  \caption{Comparison among DeepJIT, CC2Vec and {\sc CompDefect} }
  \scalebox{1}{
    \begin{tabular}{|c|c|c|c|c|c|}
    \hline
    \multicolumn{2}{|c|}{\textbf{Approach}} & \textbf{Precision} & \textbf{Recall } & \textbf{F1-score} & \textbf{AUC} \\
    \hline \hline
    \multicolumn{2}{|c|}{{DeepJIT}} & 0.504  & 0.352  & 0.414  & 0.513  \\
    \hline
    \multicolumn{2}{|c|}{{CC2Vec}} & 0.464  & 0.345  & 0.396  & 0.492  \\
    \hline
    \multicolumn{2}{|c|}{{{\sc CompDefect}}} &\textbf{0.750}  & \textbf{0.619}  & \textbf{0.679}  & \textbf{0.785}  \\
    \hline
    \multirow{2}[1]{*}{\textbf{$Improv.$}} & {$DeepJIT$} & 32.8\% & 43.2\% & 39.0\% & 34.7\% \\
\cline{2-6}          & {$CC2Vec$} & 38.2\% & 44.2\% & 41.7\% & 37.3\%  \\
\hline
    \end{tabular}%
  }
  \label{tab:rq1}
\end{table}%

\subsection{RQ2: How does {\sc CompDefect} perform on defect categorization compared with state-of-the-art baselines?}
\noindent
\textbf{Motivation.}
Even though many approaches~\cite{kamei2013large,chen2018multi,hoang2020cc2vec,liu2017code} have been proposed for just-in-time defect prediction scenario, these approaches can only predict a commit as defect-inducing or not.
They cannot categorize the type of defect.
Different from previous work, {\sc CompDefect} can categorize the type of defect the defect-inducing function belongs.
We totally consider 16 types of defects, which is introduced in Section~\ref{dataset}.
Reporting the type of defect rather than ``buggy-or-clean'' can help developers better understand such a defect.
{\sc CompDefect} makes a few progress in JIT-DP scenario even we only consider the one-statement function-level modification setting.

\noindent
\textbf{Approach.}
To verify the effectiveness of {\sc CompDefect} in defect categorization task, we choose a pre-trained language representation BERT~\cite{devlin2018bert} and its optimized variants RoBERTa~\cite{liu2019roberta} and CodeBERT~\cite{feng2020codebert}, which are widely used in natural language processing and software engineering~\cite{gao21todo,pan2021automating}, and then be used for downstream tasks that we care about (like multiclass classification).
We use these pretrained models (i.e., ``bert-base-uncased'', ``roberta-base'' and ``microsoft/codebert-base'') from Huggingface\footnote{https://huggingface.co/bert-base-uncased}.
For a fair comparison, the input of these baselines is the same as the input of {\sc CompDefect}.
That is, the buggy version function and the clean version function are treated as the input. 
In addition, since the limitation of the maximum length of baselines' input, we use the same strategy in {\sc CompDefect} to concatenate the two function vertically to assemble the whole input.
Besides, for evaluate the performance of {\sc CompDefect} and these baselines in defect categorization scenario, we also use the four performance measures (i.e., Precision, Recall, F1-score and AUC).
These classification metrics are defined for binary cases by default. 
When extending these binary metrics to multiclass, we use the ``macro'' averaging strategies, which are widely adopted in prior work~\cite{pan2021automating,arya2019analysis,wood2018detecting,huang2018automating}.
``macro'' strategy first calculates each metric for each class and then report the average value among all classes.

\noindent
\textbf{Result.}
The evaluation results are reported in Table~\ref{tab:rq2} and the best results are highlighted in bold.
On average, we find that {\sc CompDefect} outperforms 
baselines by 63\%, 218\% and  239\% in terms of  Precision, Recall, and F1-score, respectively.
In particular, {\sc CompDefect} improves BERT by 33\%, 250\% and 284\% in terms of Precision, Recall and F1-score, respectively.
Compared with RoBERTa, {\sc CompDefect} improves RoBERTa by 77\%, 245\% and 275\% in terms of Precision, Recall and F1-score, respectively, which means that even though {\sc CompDefect} and RoBERTa have the same architecture, {\sc CompDefect} can learn more information with its related task (i.e., defect fix).
Compared with CodeBERT, {\sc CompDefect} improves CodeBERT by 78\%, 158\% and 159\% in terms of Precision, Recall and F1-score, respectively, which also means {\sc CompDefect} can benefit from multi-task learning.
In terms of AUC, {\sc CompDefect} still performs best.
All the results indicate the priority of {\sc CompDefect} on categorizing the types of defects.

\begin{table}[htbp]
  \centering
  \caption{Comparison among BERT, RoBERTa, CodeBERT and {\sc CompDefect} on explaining the types of defect}
  \scalebox{1}{
    \begin{tabular}{|c|c|c|c|c|c|c|}
    \hline
    \multicolumn{2}{|c|}{\textbf{Approach}} & \textbf{Precision} & \textbf{Recall} & \textbf{F1-score} & \textbf{AUC$_{OVO}$} & \textbf{AUC$_{OVR}$} \\
    \hline \hline
    \multicolumn{2}{|c|}{BERT} & 0.301  & 0.084  & 0.083  & 0.693  & 0.731  \\
    \hline
    \multicolumn{2}{|c|}{RoBERTa} & 0.227  & 0.086  & 0.085  & 0.712  & 0.751  \\
    \hline
    \multicolumn{2}{|c|}{CodeBERT} & 0.226  & 0.115  & 0.124  & 0.695  & 0.754  \\
    \hline
    \multicolumn{2}{|c|}{\sc CompDefect} & \textbf{0.401 } & \textbf{0.295 } & \textbf{0.319 } & \textbf{0.723 } & \textbf{0.776 } \\
    \hline
    \multirow{4}[0]{*}{$Improv.$} & $BERT$  & 33\%  & 250\% & 284\% & 4\%   & 6\% \\
\cline{2-7}          & $RoBERTa$ & 77\%  & 245\% & 275\% & 2\%   & 3\% \\
\cline{2-7}          & $CodeBERT$ & 78\%  & 158\% & 159\% & 4\%   & 3\% \\
\cline{2-7}          & $Avg.$  & 63\%  & 218\% & 239\% & 3\%   & 4\% \\
\hline
\multicolumn{7}{l}{``OVO'': stands for one-vs-one; ``OVR'': stands for one-vs-rest.}\\
    \end{tabular}%
  }
  \label{tab:rq2}%
\end{table}%

\subsection{RQ3: How does {\sc CompDefect} perform on defect repair compared with the state-of-the-art baseline?}
\noindent
\textbf{Motivation.}
Defect repair research is active and mostly dominated by techniques based on static analysis~\cite{mechtaev2016angelix} and dynamic analysis~\cite{wen2018context}. 
Even a great progress has been achieved, currently, the state of automated defect repair is limited to simple cases, mostly one-line patches~\cite{wen2018context,saha2017elixir}. 
Recently, defect repair tools based on machine learning especially for deep learning technology are proposed, which promotes the further development of defect repair.
In particular, SequenceR, proposed by Chen et al.~\cite{chen2019sequencer}, is one of the most outstanding ones, which also mainly focus on one-line patch scenario.
Therefore, we want to evaluate the performance difference between {\sc CompDefect} and SequenceR.\\
\textbf{Approach.}
SequenceR can solely address one-line level bug fix. 
That is, it cannot identify whether a function is defect-inducing one and cannot categorize the type of defect in such a function.
Therefore, we filter negative functions in original dataset and keep only positive functions.
We refer to it as Function-SStuBs4J$_{\mathit{positive}}$, which contains 16 types of positive functions.
Besides, SequenceR firstly does an abstraction operation on function.
However, some function in Function-SStuBs4J$_{\mathit{positive}}$ cannot be executed  successfully with the tool provided by SequenceR.
Thus, we filter out these functions and finally Function-SStuBs4J$_{\mathit{positive}}$ has the positive function with 14 types.
{We split 80\%, 10\% and 10\% of Function-SStuBs4J$_{\mathit{positive}}$ as training data (7,474 functions), validation data (934 functions) and testing data (934 functions), respectively}, and the distribution among each type of function as same as the original one.
For a fair comparison, we train SequenceR and {\sc CompDefect} ({referred as {\sc CompDefect}$_\mathit{positive}$}) on the filtered training data, optimize them on the filtered validation data, and finally evaluate them on the filtered testing dataset.

Moreover, for fully evaluating the capability of {\sc CompDefect}, we want to directly evaluate the {\sc CompDefect} trained on original training data (i.e., training data from Function-SStuBs4J, denoted as Training$_{\mathit{original}}$) on the testing data of Function-SStuBs4J$_{\mathit{positive}}$, denoted as Testing$_{\mathit{positive}}$.
However, since Function-SStuBs4J$_{\mathit{positive}}$ and Function-SStuBs4J are not exactly the same split, Training$_{\mathit{original}}$ may contains the functions in Testing$_{\mathit{positive}}$.
So, for a fair comparison, we identify the intersection of Training$_{\mathit{original}}$ and Testing$_{\mathit{positive}}$, and remove the intersection from Testing$_{\mathit{positive}}$ and denote it as Testing$_{\mathit{positive}\_\mathit{filtered}}$.
For evaluating the two methods, we adopt two widely used performance measures: BLEU and Accuracy.
\newline\noindent
\textbf{Result.}
{
The evaluation results are reported in Table~\ref{tab:rq3} and the best results are highlighted in bold.
On Testing$_{positive}$, {\sc CompDefect}$_\mathit{positive}$ performs better than SequenceR and outperforms SequenceR by 23.9\% and 29.5\% in terms of BLEU and Accuracy, respectively.
On Testing$_{positive\_filtered}$, {\sc CompDefect}$_\mathit{positive}$ can also improve SequenceR by 29.6\% in terms of BLEU and perform similarly as SequenceR in terms of Accuracy.
{\sc CompDefect} also outperform SequenceR on both performance measure.
Besides, when comparing {\sc CompDefect} with {\sc CompDefect}$_\mathit{positive}$, we find that {\sc CompDefect} has better performance, which indicates that {\sc CompDefect} benefits from the multi-task learning and learns useful information from negative functions for defect repair.
}

\begin{table}[htbp]
  \centering
  \caption{Comparison between SequenceR and {\sc CompDefect} on defect repair }
  \scalebox{1}{
    \begin{tabular}{|c|c|c|c|}
    \hline
    \textbf{Testing Data} & \textbf{Approach} & \textbf{BLEU} & \textbf{Accuracy} \\
    \hline \hline
    \multirow{2}[0]{*}{{Testing$_{positive}$}} & {SequenceR} & 0.734 & 0.136 \\
\cline{2-4}          & \textbf{{\sc CompDefect}$_\mathit{positive}$} & \textbf{0.964} & \textbf{0.193} \\
\hline \hline
    \multirow{3}[0]{*}{{Testing$_{positive\_filtered}$}} & {SequenceR} & 0.677 & 0.147 \\
\cline{2-4}          & \textbf{{\sc CompDefect}$_\mathit{positive}$} & 0.962 & 0.130 \\
\cline{2-4}          & {{\sc CompDefect}} & \textbf{0.972} & \textbf{0.232} \\
\hline
    \end{tabular}%
  }
  \label{tab:rq3}%
\end{table}%


\vspace{-.5cm}
\section{Threats to Validation}
\label{threats}

\textbf{Threats to Internal Validity} mainly lie in the potential faults in the implementation of our model. 
To minimize such threats, we not only implement these methods by pair programming but also make full use of the pre-trained models such as GraphCodeBERT~\cite{guo2020graphcodebert} and BERT~\cite{devlin2018bert}.
Besides, we directly use the original source code of baselines and the same hyperparameters used in original method are adopted in our paper. 
All of the datasets used in our study are publicly available
from previous work~\cite{karampatsis2020often}, and we extend these dataset for our investigated scenario.

\noindent
\textbf{Threats to External Validity} mainly lie in the studied projects used in this study. To reduce
such threats, we opted to selecting high popularity Java projects.
The popularity of a project is determined by computing the sum of $z$-scores of forks and stars~\cite{karampatsis2020often}. 
However, all studied projects are open source projects, it is still unknown whether our {\sc CompDefect} can work well on commercial projects. 
In the future, we plan to reduce this threat by considering more additional commercial projects.

\noindent
\textbf{Threats to Construct Validity} mainly lie in the adopted performance metrics in our evaluations. To reduce such threat, we use different types of performance measure for different tasks.
For classification task, we use the widely-used Area Under the Curve (AUC) score to evaluate the performance of the each method and it needs to be set a threshold manually. 
We also consider some widely performance metrics which need to manually set a threshold e.g., Precision, Recall and F1-score.
For generation task, we use two well-known metrics namely BLEU and Accuracy.

\section{Related Work}
\label{relatedwork}

\subsection{Just-in-Time Defect Prediction}

    JIT defect prediction has been an active research topic in recent years since it can identify defect-inducing commit at a fine-grained level at check-in time.
    Mockus et al.~\cite{mockus2000predicting} extracted historical information in commit to build a classifier to predict the risk of new commits.
    Kamei et al.~\cite{kamei2013large} proposed 14 change-level metrics from five dimensions, which are used to build an effort-aware prediction model with the help of Logistic Regression.
    Following that, Yang et al.~\cite{yang2015deep,yang2017tlel} subsequently proposed two approach for JIT defect prediction. 
    In particular, Yang et al.~\cite{yang2015deep} firstly used Deep Belief Network (DBN) to extract higher-level information from the initial change-level metrics, then Yang et al.~\cite{yang2017tlel} combined decision tree and ensemble learning to build a two-layer ensemble learning model for JIT defect prediction.
    To further improve Yang et al's model, Young et al.~\cite{young2018replication}
    proposed a new deep ensemble approach by using arbitrary classifiers in the ensemble
    and optimizing the weights of the classifiers.
    Later, Liu et al.~\cite{liu2017code} proposed code churn and evaluated it in effort-aware settings. 
    Chen et al.~\cite{chen2018multi} treated the effort-aware JIT defect prediction task as a multi-objective optimization problem and consequently a set of effective features are selected to build the prediction model.
    Then, Cabral et al.~\cite{cabral2019class} proposed a new sampling technology to address the issues of verification latency and class imbalance evolution in online JIT defect prediction setting. 
    Besides, Yan et al.~\cite{yan2020just} proposed a two-phase framework which can handle the identification and localization tasks at the same time.
    Recently, Hoang et al.~\cite{hoang2019deepjit,hoang2020cc2vec} proposed two newly approaches, which use modern deep learning model to learn the representation of commit message and code changes.

Apart from above approaches, researchers also conduct studies on JIT defect prediction from various aspects.
McIntosh et al.~\cite{McIntoshK18} investigated the impact of systems evolution on JIT defect prediction models via a longitudinal case study of 37,524 changes from the rapidly evolving QT and OpenStack systems. 
They found that the interval between training periods and testing periods has side-effect on the performance of JIT models and JIT models should be trained using six months (or more) of historical data.
Besides, Wan et al.~\cite{wan2018perceptions} discussed the drawbacks of existing defect prediction tools and highlighted future research directions through literature review and a survey of practitioners.
After that, 
Tabassum et al.~\cite{tabassum2020investigation} conducted a study of JIT defect prediction in realistic online learning scenarios and concluded that the model trained with both within and
cross-project data can outperform the model trained with within-project data only. 
Recently, Zeng et al.~\cite{Zeng2021Deep} revisited the deep learning based JIT defect prediction models and found that deep learning based approaches may not work better than simplistic model LApredict they proposed.

Different from the existing work (i.e., commit-level JIT defect prediction), our model {\sc CompDefect} focus on function-level single-statement JIT defect prediction, which is more fine-grained defect identification task.
Besides, previous work only give two coarse-grained outputs: defect-inducing or clean.
{\sc CompDefect} can categorize the type of defect-inducing functions.

\subsection{Defect Repair}

Defect repair~\cite{gazzola2017automatic} is an active research topic and achieve a great progress.
However, the current state of automated defect repair is limited to simple small fixes, mostly one-line
patches~\cite{wen2018context,saha2017elixir}. 
Gupta et al.~\cite{gupta2017deepfix} proposed a defect repair tool named DeepFix for fixing compiler
errors in introductory programming courses. 
DeepFix takes the whole program and outputs a single line fix.
Ahmed et al.~\cite{ahmed2018compilation} also focused on compiler error fix and proposed another defect repair tool named TRACER, which outperforms DeepFix in terms of success rate. 
Martin et al.~\cite{white2019sorting} proposed DeepRepair to leverage the learned code similarities to select repair ingredients from code fragments which are similar to the buggy code. 
Tufano et al.~\cite{tufano2019empirical} performed an empirical study to assess the feasibility of using neural machine translation techniques for learning bug-fixing patches for real defects.
They trained an Encoder-Decoder model which can translate buggy code into its fixed version.
Hideaki et al.~\cite{hata2018learning} proposed a similar network and applied it to one-line diffs.

Considering the complexity of defect repair,  similar to previous work, our model {\sc CompDefect} also focus on the one-line code fix scenario.
However, different from the existing work, the input of {\sc CompDefect} is the source code of changed function in a commit and it aims to provide a foundation for connecting defect identification and defect repair.
{\sc CompDefect} can categorize the type of a defect and can fix it at the check-in time.

\subsection{Pre-Trained Models in NLP and SE}

Pre-trained technologies have achieved a big success in Natural Language Processing (NLP)~\cite{devlin2018bert,yang2019xlnet,liu2019roberta} and pre-trained models associated with programming languages~\cite{feng2020codebert,karampatsis2020scelmo,lewis2019bart} also make a great process of code intelligence.
Kanade et al.~\cite{kanade2019pre} pre-trained a BERT model on a massive corpus of Python source codes through two tasks: masked language modeling and next sentence prediction. 
Lewis et al.~\cite{lewis2019bart} adopted a standard Transformer-based neural network to train a sequence-to-sequence model named BART.
Later, Feng et al.~\cite{feng2020codebert} used Transformer-based neural architecture to propose a
bimodal pre-trained model named CodeBERT for programming language and natural language by masked language modeling and replaced token detection.
CodeBERT aims at learning general-purpose representations for supporting downstream NL-PL applications (e.g., code search, code documentation generation).
Karampatsis et al.~\cite{karampatsis2020scelmo} pre-trained contextual embeddings on a JavaScript corpus by using the ELMo frame, which can be further used for program repair task. 
Guo et al.~\cite{guo2020graphcodebert} proposed a pre-trained model named GraphCodeBERT which can leverage code structure to learn code representation to improve code understanding.

Recently, pre-trained models are widely used in software engineering (SE) and these models are used to learn the representation of source code.
Gao et al.~\cite{gao21todo} proposed an approach named TDCleaner, a BERT-based neural network, to automatically detect and remove obsolete TODO comments from software repositories. 
In this paper, we also use GraphCodeBERT to generate the representation of function.

\section{Conclusion and Future Work}
\label{conclusion}

In this paper, we propose a comprehensive  defect prediction and repair framework named {\sc CompDefect}, which can identify whether a function changed in a commit is defect-prone, categorize the  type of  defect, and repair such a defect automatically.
Generally, the first two tasks in {\sc CompDefect} is treated as a multiclass classification task, while the last one is treated as a sequence generation task.
The whole input of {\sc CompDefect} consists of three parts: the clean version of function (i.e., the version before the defect introduced), the buggy version of function and the fixed version of function.
For the first task, {\sc CompDefect} identifies the defect type through multiclass classification.
For the second task, {\sc CompDefect} repairs the defect once identified or keeps it as the same originally.
Moreover, we also build a new function-level dataset on the basis of ManySStuBs4J to evaluate the performance of {\sc CompDefect}.
The new dataset is the largest function-level dataset with comprehensive information.
It contains three versions of a certain function and multiple types of defect.
By comparing with state-of-the-art baselines in various settings, {\sc CompDefect} can achieve superior performance on classification and defect repair.

{
    In the future, we firstly want to make our approach an online available tool for practical usage.
    Then, we will extract more non-single-statement defective functions and further improve our model to address this situation.
}

\bibliographystyle{ACM-Reference-Format}
\bibliography{CompDefect}
\end{document}